\documentclass[aip,jcp,reprint]{revtex4-1}
\usepackage{amsmath}
\usepackage{graphicx}
\usepackage{hyperref}
\usepackage[capitalise]{cleveref}
\usepackage{braket}

\usepackage{color}

\begin{document}
\title{Linked Coupled Cluster Monte Carlo}
\author{R. S. T. Franklin}
\affiliation{University Chemical Laboratory, Lensfield Road, Cambridge, CB2 1EW, United Kingdom}
\author{J. S. Spencer}
\affiliation{Department of Physics, Imperial College London, Exhibition Road, London, SW7 2AZ, United Kingdom}
\affiliation{Department of Materials, Imperial College London, Exhibition Road, London, SW7 2AZ, United Kingdom}
\author{A. Zoccante}
\affiliation{University Chemical Laboratory, Lensfield Road, Cambridge, CB2 1EW, United Kingdom}
\author{A. J. W. Thom}
\affiliation{University Chemical Laboratory, Lensfield Road, Cambridge, CB2 1EW, United Kingdom}
\date{\today}
\begin{abstract}
    We consider a new formulation of the stochastic coupled cluster method in terms of the similarity transformed Hamiltonian.
    We show that improvement in the granularity with which the wavefunction is represented results in a reduction in the critical population required to correctly sample the wavefunction for a range of systems and excitation levels and hence leads to a substantial reduction in the computational cost.
    This development has the potential to substantially extend the range of the method, enabling it to be used to treat larger systems with excitation levels not easily accessible with conventional deterministic methods.
\end{abstract}


\maketitle 

Coupled cluster theory is a highly accurate and size consistent method of treating electron correlation.\cite{Cizek66,Cizek71} It can achieve high accuracy for molecules \cite{Bartlett07} as well as for extended systems \cite{Gruneis11, Gruneis15}.
However, the scaling with system size ($O(N^6)$ for the CCSD truncation and worsening for higher truncation) and the complexity of the equations at high truncation levels limit the applicability of the method.
It also scales poorly over distributed computer architectures due to the dense linear algebra operations involved.
Recently, inspired by the successes of the Full Configuration Interaction Quantum Monte Carlo (FCIQMC) method \cite{Booth09,Booth10,Booth11,Cleland12,Gruneis11,Booth13,Booth12,Overy14,Shepherd14}, a stochastic formulation of the coupled cluster equations (coupled cluster Monte Carlo, CCMC) has been proposed to ameliorate these problems. \cite{Thom10}
It uses a set of discrete ``excips'' to represent the cluster amplitudes, but only converges to the correct ground state when a system dependent critical number of walkers (the ``plateau'') is exceeded.\cite{Thom10,Spencer12}
This limits the size of systems accessible to the method.
In this paper, we propose a modification of the CCMC algorithm that reduces the plateau heights enabling the application of the method to larger systems.

In coupled cluster theory, the wavefunction is represented using the exponential ansatz $\Psi_\mathrm{CC} = \mathrm e^{\hat{T}} \Ket{ D_\mathrm{HF} }$, where $\hat{T} = \sum_\mathbf{i} t_\mathbf{i} \hat{a}_\mathbf{i}$ and $\hat{a}_\mathbf{i}$ is a string of creation and annihilation operators in second quantization theory, such that $
\hat{a}_\mathbf{i} \Ket{ D_\mathrm{HF} } = \Ket{ D_\mathbf{i} }$, where $\Ket{ D_\mathbf{i} }$ is the Slater determinant with orbitals $i_1, i_2, \ldots, i_N$ occupied.
Truncations of the theory are obtained by restricting the sum in $\hat{T}$ to contain excitations of at most the truncation level number of orbitals, but the exponential form ensures the theory remains size consistent.
The amplitudes $t_\mathbf{i}$ are determined by projecting the Schr\"odinger equation $\hat{H} \Psi_\mathrm{CC} = E \Psi_\mathrm{CC}$ onto the set of substituted determinants generated by $\hat{T}$.
In CCMC, after rescaling the amplitudes by $t_\mathrm{HF}$ to give the modified ansatz $\Psi_\mathrm{CC} = t_\mathrm{HF} \mathrm{e}^{\hat{T}/t_\mathrm{HF}} \Ket{ D_\mathrm{HF} }$, this is recast into an iterative form:
\begin{equation}
    \label{eq:CCMC}
    t_\mathbf{i} \rightarrow t_\mathbf{i} - \delta\tau\Braket{ D_\mathbf{i} | \hat{H} - S | \Psi_\mathrm{CC} }
\end{equation}
 where $\delta\tau$ is a small positive number (the timestep) and $S$ the shift, an energy offset introduced to control the growth of the coefficients.
The coefficients of the wavefunction are discretised and represented by populations of fictitious particles called excips on each excitor, where an excitor is a combination of creation and annihilation operators that produces an excited determinant from the Hartree-Fock.
The wavefunction and Hamiltonian are then stochastically sampled to update the excitation amplitudes.
In outline, the algorithm proceeds as follows (for more details see Ref.~\onlinecite{SpencerThom15}): \\

\begin{enumerate}
    \item Choosing a cluster: $s$ excitors ($s$ is randomly chosen from $ 0 \leq s \leq l+2$ where $l$ is the truncation level) are randomly chosen and combined to give an excitation operator which acts on the reference determinant to give a determinant $\Ket{ D_\mathbf{n} }$.
        This step is equivalent to truncation of the Taylor expansion of $e^{\hat{T}}$ at $l+2$-th power (exact for a 2-body Hamiltonian with up to $l$-fold excitations in $\hat{T}$) and samples the wavefunction ansatz.
        The factor of $\frac{1}{n!}$ arising in the Taylor expansion is absorbed into the calculated weight of the cluster.
    \item Spawning: A random single or double excitation of $\Ket{ D_\mathbf{n} }$, $\Ket{ D_\mathbf{m} }$, is chosen and a new excip spawned on $\Ket{ D_\mathbf{m} }$ with probability proportional to $|H_\mathbf {mn}|$ and sign determined by the sign of $H_\mathbf {mn}$, where $H_\mathbf{mn} = \braket{ D_\mathbf{m} | \hat{H} | D_\mathbf{n}}$.  This step samples the action of the Hamiltonian.
    \item Death: A new excip is spawned on $\Ket{ D_\mathbf{n} }$ with probability proportional to $|H_\mathbf {nn} - E|$ and opposite sign to the parent.
    \item Annihilation: Pairs of excips with opposite signs on the same excitor are removed.
\end{enumerate}

In conventional treatments of coupled cluster theory, it is convenient to rewrite the equations in the form
\begin{equation}
    \Braket{ D_\mathbf{m} | \bar{H} - E | D_\mathrm{HF} } = 0,
\end{equation}
 using the similarity transformed Hamiltonian $\bar{H}~=~e^{-\hat{T}} \hat{H} e^{\hat{T}}$.
 This has the advantage that, using the Campbell--Baker--Hausdorff formula and Wick's theorem\cite{Mattuck}, the expansion of the exponential truncates at fourth order and can be written as 
\begin{align}
    \bar{H} = \hat{H} &+ [\hat{H}, \hat{T}]_c + \frac{1}{2} [[\hat{H}, \hat{T}], \hat{T}]_c + \frac{1}{3!} [[[\hat{H}, \hat{T}], \hat{T}], \hat{T}]_c \nonumber\\
                        &+  \frac{1}{4!} [[[[\hat{H}, \hat{T}], \hat{T}], \hat{T}], \hat{T}]_c,\label{eq:CBH}
\end{align}
where the subscript $c$ is included to emphasise that only the terms of the commutators in which the Hamiltonian is connected to each cluster operator are needed, and the remaining terms cancel to zero due to the commutators.
This decouples the amplitude equations from the energy equation.

The question thus arises as to whether benefits can be derived from using the similarity transformed Hamiltonian in CCMC.
The iterative equations used to calculate the amplitudes are, analogously to the original CCMC equations,
\begin{align}
    \label{eq:lCCMC}
    t_\mathbf{i} &\rightarrow t_\mathbf{i} - \delta\tau t_\mathrm{HF} \Braket{ D_\mathbf{i} | \bar{H} | D_\mathrm{HF} } \qquad \text{ if } \Ket{ D_\mathbf{i} } \neq \Ket{ D_\mathrm{HF} } \\
 t_\mathrm{HF} &\rightarrow t_\mathrm{HF} - \delta\tau t_\mathrm{HF} \Braket{ D_\mathrm{HF} | \bar{H} - S | D_\mathrm{HF} }
\end{align}
Notably, the shift only appears in the equation for the reference population. This means that the population control only acts directly on the reference, and population control for other excitors is achieved indirectly by changes in the rate of spawning, both from the reference and also, due to the change in normalisation, from the other excitors.

Can we sample $\bar{H}$ in a manner analogous to CCMC?
In \cref{eq:CBH}, we may choose any of the five terms at random, and this is equivalent to selecting the size, $s$, of cluster to consider.
Each $\hat{T}$ within the unexpanded commutator is itself a sum over excitors (each with an amplitude).
To sample this, for each of the $\hat{T}$ operators, we pick, at random, just one of the excitors within its sum.
This results in $s$ (potentially different) excitors, i.e. a cluster of size $s$.

Given this specific choice of excitors, we must still evaluate the entire nested commutator.
Could we instead sample at random just one of the terms in the expansion of the nested commutator rather than the complete expansion?
While this is possible, we note that one of the benefits of the similarity transformed approach is that for unlinked clusters all the terms in the nested commutator cancel to zero, and so decomposing the commutator would remove any gains from the cancellation of terms.
In order to sample the commutator we must instead evaluate the sum of all the permutations of $\hat{H}$ and $\hat{T}$ operators arising from the commutators for a sampled term from $\bar{H}$.
The random sampling of a cluster containing $s$ excitors from the $\hat{T}^s$ term in the power expansion of $\mathrm{e}^{\hat{T}}$ in conventional CCMC can also be used to select the excitors included in the $s$-fold commutator.
However, the allowed excitations in $\hat{H}$ (to give a non-zero projection on to some $D_\mathbf{i}$ in \cref{eq:lCCMC}) can differ in each term within the commutator, and so sampling this is not so straightforward.

To solve this, we return briefly to look at the second sampling step in CCMC: $\hat{H}$ is sampled by selecting an allowed excitation of the collapsed cluster $\Ket{ D_\mathbf{n} }$.  i.e. we choose $\Bra{ D_\mathbf{i}}$ such that matrix element $H_\mathbf{in}=\Bra{ D_\mathbf{i}}\hat{H}\Ket{D_\mathbf{n}}$ is non-zero.
Once this excitation is chosen, the value of $\mathbf i$ has been determined, and this then selects a single $t$-update, $t_\mathbf i$, from the set of all possible updates in \cref{eq:CCMC}.
We may look at this in a different manner:  Let us select a given projectee determinant $\Bra{ D_\mathbf{i} }$ in \cref{eq:CCMC}, and see what allowed samplings of $\hat{H}$ may lead to it.
For conventional CCMC, the choice is unique.

For linked CCMC, a choice of a given projectee determinant $\Bra{ D_\mathbf{i} }$  corresponds to at most one allowed excitation in $\hat{H}$ in each term of the commutator.  This excitation is the same in each term, but may have a different corresponding matrix element.
Therefore for a given sampling of $\hat{T}^s$ (i.e. cluster), we must carefully select one from all possible allowed projectees, and then we may evaluate the commutator for that projectee.

To use this form of the coupled cluster equations in stochastic coupled cluster requires making three modifications to the steps in conventional CCMC. 
\begin{enumerate}
    \item Due to the truncation of the Hausdorff expansion, only clusters of at most four excitors need to be used, regardless of the truncation level considered.
    \item Clusters where two of the operators excite to or from the same orbital (following Bartlett et. al.\cite{Bartlett1978}, we refer to these as ``conjoint'' terms, though they are also known as Exclusion-Principle-Violating terms\cite{ShavittBartlett}) must also be considered.
          For example, take a pair of excitors that excite from the same orbital, say $\hat{a}_i^a$ and $\hat{a}_i^b$, where $\hat{a}_i^a=\hat{c}_i\hat{c}_a^{\dagger}$ and $\hat{c}_i$ ($\hat{c}_a^{\dagger}$) is a conventional annihilation (creation) operator.
          $\hat{a}_i^a \hat{a}_i^b = 0$ and hence $\braket{ D_\mathbf{i} | \hat{H} \hat{a}_i^a \hat{a}_i^b | D_\mathrm{HF} } = 0$ for all $\Ket{ D_\mathbf{i} }$ and in the original form of stochastic coupled cluster any such cluster of excitors can be ignored.  The argument applies equally to a pair of excitors that excite to the same orbital.

          When the similarity transformed Hamiltonian is used, such a cluster can give a non-zero contribution, however, as terms from the commutators, such as $\braket{ D_\mathbf{i} | \hat{a}_i^a \hat{H} \hat{a}_i^b | D_\mathrm{HF} }$ are not necessarily 0.
          This means that the excitation in $\hat{H}$ must be chosen from the determinant produced by collapsing some subset of the cluster, $\Ket{ D_\mathbf {n} }$, here given by $\hat{a}_i^b \Ket{ D_\mathrm{HF} } = \Ket{ D_i^b }$, as collapsing the whole cluster just gives zero, not a valid excitor.
          From this, excitation $\Bra{ \mathbf{m} }$ is chosen with non-zero $H_\mathbf{mn}$ to sample $\hat{H}$.
          The projectee $\Bra{ D_\mathbf{i} }$ follows from applying the remaining excitors to $\Ket{ D_\mathbf{m} }$.
          This overall process corresponds to the product $\braket{ D_\mathbf{i} | \hat{a}_i^a | D_\mathbf{m} } \braket{ D_\mathbf{m} | \hat{H} | D_\mathbf{n} } \braket{ D_\mathbf{n} | \hat{a}_i^b | D_\mathrm{HF} }$

        Different partitions may give non-zero contributions to the commutator for different projectees, so to sample all possible spawning events in an unbiased manner requires randomly selecting a partition, using that to choose an allowed projectee, then evaluating the commutator by considering all possible (allowed) partitions.  

    \item All Hamiltonian matrix elements, in spawning and death, must be replaced by matrix elements of the similarity transformed Hamiltonian.
          This also means that the shift only applies to death on the reference, and there is no death for conjoint clusters.
          For instance when a cluster formed of a single excitor $\hat{a}_i^a$ is being considered and the excitation for the Hamiltonian chosen for spawning is $\hat{a}_j^b$, the probability of spawning an excip on $\Ket{ D_{ij}^{ab} }$ does not depend on $H_\mathbf{mn} = \braket{ D_{ij}^{ab} | \hat{H} | D_i^a }$ but on $\braket{ D_\mathbf{i} | [\hat{H}, \hat{T}] | D_\mathrm{HF} } = \braket{ D_{ij}^{ab} | \hat{H} | D_i^a } - \braket{ D_{j}^{b} | \hat{H} | D_\mathrm{HF} }$. 
\end{enumerate}

Consider a system of three occupied orbitals, $i$, $j$ and $k$, and three virtual orbitals, $a$, $b$ and $c$. Some representative spawning attempts are:
\begin{enumerate}
    \item A cluster consisting of a single excitor $\hat{a}_i^b$ is selected, giving the determinant $\Ket{ D_i^b }$ when applied to the reference.
        The single excitation to $\Bra{ D_{ij}^{ab} }$ is chosen for the spawning attempt, which succeeds with probability proportional to $|\braket{ D_{ij}^{ab} | [\hat{H}, \hat{a}_i^b ]| D_\mathrm{HF} }| = |\braket{ D_{ij}^{ab} | \hat{H} | D_i^b } - \braket{ D_j^a | \hat{H} | D_\mathrm{HF} }|$.
        Death also occurs on $D_i^b$ with probability proportional to $\braket{ D_i^b | \hat{H} | D_i^b } - \braket{ D_\mathrm{HF} | \hat{H} | D_\mathrm{HF} }$.

    \item The cluster composed of the two excitors $\hat{a}_i^a$ and $\hat{a}_j^b$ is chosen, and collapsed onto the reference to give $\Ket{ D_{ij}^{ab} }$.
        The single excitation to $\Ket{ D_{ijk}^{abc} }$ is chosen for spawning, but this excitation ($\hat{a}_k^c$) is not linked to the cluster so no spawning is attempted as all the terms in the commutator cancel.
        Death occurs on $\Ket{ D_{ij}^{ab} }$ with probability proportional to $\braket{ D_{ij}^{ab} | \hat{H} | D_{ij}^{ab} } - \braket{ D_i^a | \hat{H} | D_i^a } - \braket{ D_j^b | \hat{H} | D_j^b } + \braket{ D_\mathrm{HF} | \hat{H} | D_\mathrm{HF} }$.

    \item The cluster composed of two excitors $\hat{a}_i^a$ and $\hat{a}_i^b$ is chosen.
          These both excite from the same occupied orbital $i$ so do not collapse to give an excitor that can be applied to the reference.
          To choose an excitation for the spawning step, the subset of the cluster which does produce a determinant must be chosen.
          The subset containing only $\hat{a}_i^a$ is randomly chosen to be the latter, corresponding to the partitioning $\braket{ D_\mathbf{i} | \hat{a}_i^b \hat{H} \hat{a}_i^a | D_\mathrm{HF} }$, giving the determinant $\Ket{ D_i^a }$ to spawn from.
          The double excitation ($\hat{a}_{aj}^{ic}$) to $\Ket{ D_j^c }$ is chosen in the Hamiltonian, and then the remainder of the original cluster $\hat{a}_i^b$ is applied to give spawning on $\Ket{ D_{ij}^{bc} }$ with probability proportional to $|\braket{ D_j^c | \hat{H} | D_i^a }|$.
          There is no death attempt for conjoint clusters.
\end{enumerate}

\begin{figure}[ht]
    \centering
    \includegraphics[width=\columnwidth]{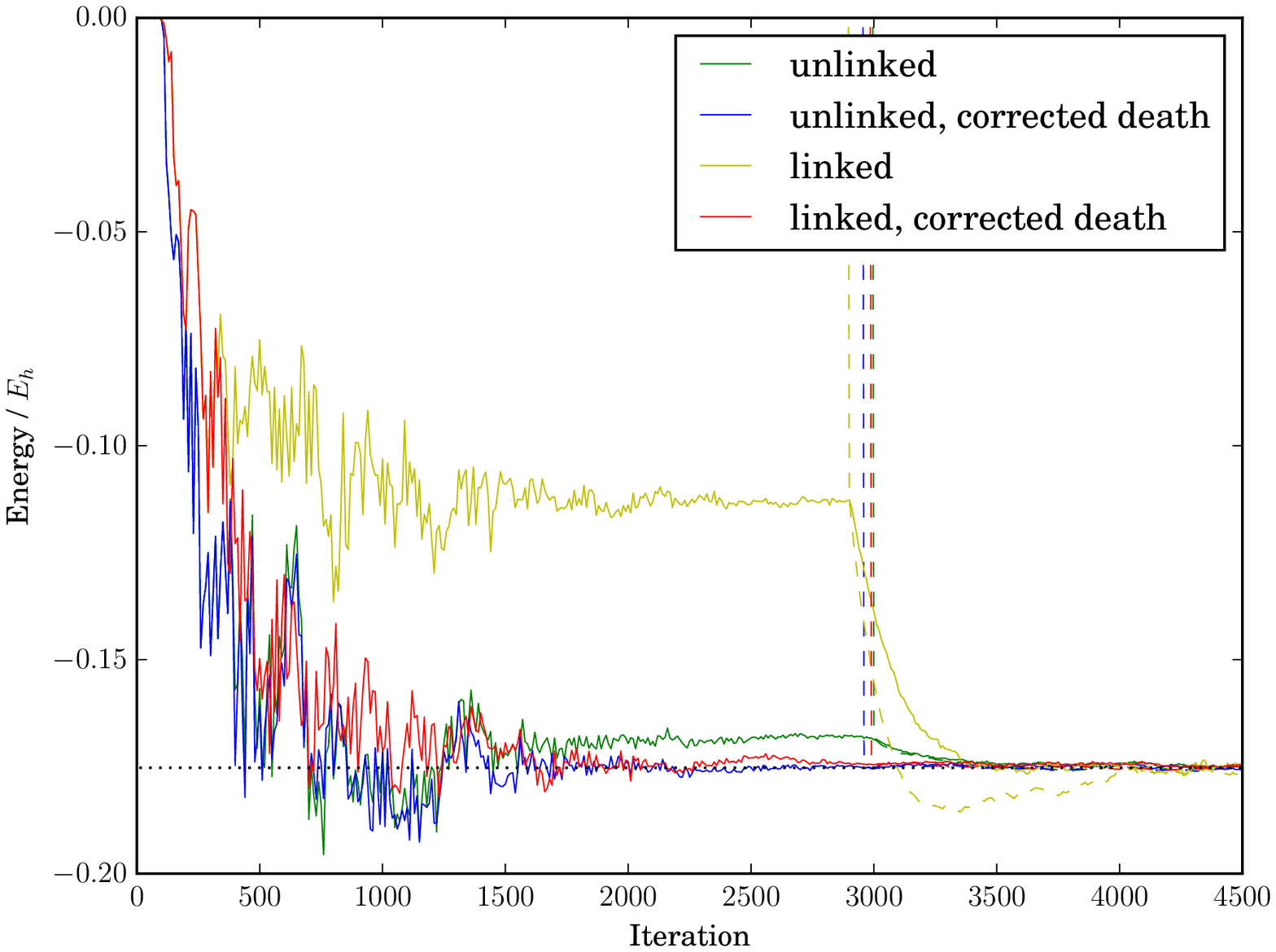}
    \caption{The correlation energy calculated in CCMC for the pathological LiH$_2^+$ system\footnote{This was generated for minimal debugging purposes and has geometry $r_{\mathrm{LiH}_1}=1$\,\AA, $r_{\mathrm{LiH}_2}=1.1$\,\AA, and $\theta_{\mathrm{HLiH}}=90\,^\circ$. In the STO-3G basis, unrestricted orbitals were prepared with a custom density functional with purely exchange, $E_{\mathrm x}= 2K + X_{\mathrm{Slater}} + X_{\mathrm{Becke88}}$. 2 core electrons and 8 virtual spinorbitals were frozen leaving an active space of 2 electrons in 4 spinorbitals. Q-Chem outputs and molecular orbital integrals for this system are included in the data repository.\cite{linked_CC_data}} at the CCSD level; the projected energy is given by solid lines, the shift by dashed lines and the exact CCSD energy ($-0.175277 \mathrm{E}_h$ ) by the black dotted line.  An incorrect projected energy is obtained in the fixed-shift regime using the simple form of death.  The shift and projected estimators fluctuate about the exact CCSD energy in the variable-shift regime in all cases.\label{fig:CCMC-shift}}
\end{figure}

When the algorithm described above is used to calculate the coupled cluster energy, convergence to the correct projected energy is only achieved in the variable-shift regime, as shown in \cref{fig:CCMC-shift}.  The system shown in \cref{fig:CCMC-shift} is a pathological system displaying a very pronounced bias but the same behaviour is general to all systems examined, albeit to a lesser degree.  Similar behaviour is observed in FCIQMC in a non-orthogonal basis\cite{SDSThesis}.  This can be understood by considering the change in the coefficients when the projector is applied to the correct wavefunction.  The desired form once the wavefunction has converged is $\Delta t_\mathbf{i} \propto t_\mathbf{i}$, as this ensures that the wavefunction being represented does not change, only the normalisation.  Substituting the exact wavefunction into the CCMC update step gives for unlinked CCMC
\begin{align}
    \Delta t_\mathbf{i} &= -\delta\tau t_\mathrm{HF} \Braket{ D_\mathbf{i} | \hat{H}-S | \Psi_\mathrm{CC} } \\ 
                        &= -\delta\tau (E_\mathrm{exact} - S) \tilde{t}_\mathbf{i},
\end{align}
where $\tilde{t}_\mathbf{i} = \langle D_\mathbf{i} | \Psi_\mathrm{CC} \rangle$ is the coefficient of $\Ket{ D_\mathbf{i}}$ in the coupled cluster wavefunction, and for linked CCMC
\begin{align}
    \Delta t_\mathbf{i} &= -\delta\tau t_\mathrm{HF} \Braket{ D_\mathbf{i} | \bar{H} - S | D_\mathrm{HF} } \\
                        &= \begin{cases} -\delta\tau t_\mathrm{HF} (E_\mathrm{exact} - S) & \Ket{ D_\mathbf{i} } = \Ket{ D_\mathrm{HF} } \\ 0 & \text{otherwise} \end{cases}
\end{align}

For the original equations, the discrepancy $\tilde{t}_\mathbf{i} - t_\mathbf{i}$ is typically small as the connected amplitudes are the dominant contribution to the wavefunction.  However there is a much larger discrepancy when using the similarity transformed Hamiltonian, giving rise to the larger deviation in the projected energy shown in \cref{fig:CCMC-shift}.
We emphasise that this is only true in the fixed-shift regime; once the population has grown to its desired size, the shift is varied to control the population and averages to the exact energy and hence the requirement on change in coefficients is trivially met on average.

This can be resolved by modifying the update step \cref{eq:CCMC,eq:lCCMC} to be 
\begin{equation}
    t_\mathbf{i} \rightarrow t_\mathbf{i} - \delta\tau  t_\mathrm{HF} \Braket{ D_\mathbf{i} | \hat{H} - E_\mathrm{CC} | \Psi_\mathrm{CC} } - \delta \tau (E_\mathrm{CC} - S) t_\mathbf{i}, \label{eq:mod_death}
\end{equation}
and for linked CCMC
\begin{equation}
    t_\mathbf{i} \rightarrow t_\mathbf{i} -\delta\tau t_\mathrm{HF} \Braket{ D_\mathbf{i} | \bar{H} - E_\mathrm{CC}| D_\mathrm{HF} } -\delta \tau (E_\mathrm{CC} - S) t_\mathbf{i}, \label{eq:l-mod_death}
\end{equation}
both of which give $\Delta t_\mathbf{i} = -\delta\tau(E_\mathrm{CC} - S)t_\mathbf{i}$ when the $t_\mathbf{i}$ represent the exact wavefunction.

When $S = E_\mathrm{CC}$, these equations reduce to the original form, so in the variable shift regime the same solution is obtained.  The difficulty with using the \cref{eq:mod_death} and \cref{eq:l-mod_death} as written is that the exact coupled cluster energy is not known during the calculation, as it is the quantity we aim to obtain.  We therefore use the instantaneous value of the projected energy in its place as it is an easily available estimator.  This does not make a difference during the variable shift phase of the calculation, during which statistics are accumulated, but reduces the equilibration time necessary.

This two term form of the propagation can be regarded as effecting a separation between two roles of the dynamics: the first term ensures the excip distribution represents a solution to the coupled cluster equations, while the second allows control of the total population via varying the shift.

The changes to the equations amount in practice to using the projected energy rather than the shift to control the rate of death on the composite clusters.  This can be rationalised by noting that no population resides directly on the composite clusters, so such death processes do not contribute directly to population control, thus should not be dependent on the shift.

The population dynamics of the CCMC algorithm are similar to those of FCIQMC.\cite{Spencer12}
In particular, there is a plateau phase during the initial growth in the number of particles.
This plateau occurs at a system-dependent number of excips and is the manifestation of the sign problem in CCMC.
During the plateau phase, the total excip population remains constant, but the excip population on the reference continues to grow, so the plateau can be identified by finding the maximum of the ratio of the total to the reference population.\cite{SpencerThom15}
In this work we adopt the definition of the plateau height as the average population across the ten times with the largest ratio of the total population to the reference population.  We have found this to be an effective measure of the upper bound to the plateau in these systems, though note that it is somewhat dependent on the initial populations and timesteps\cite{VigorThom_arxiv}, so we have kept these the same in the calculations we have compared.

\begin{figure}[ht]
    \centering
    \includegraphics[width=\columnwidth]{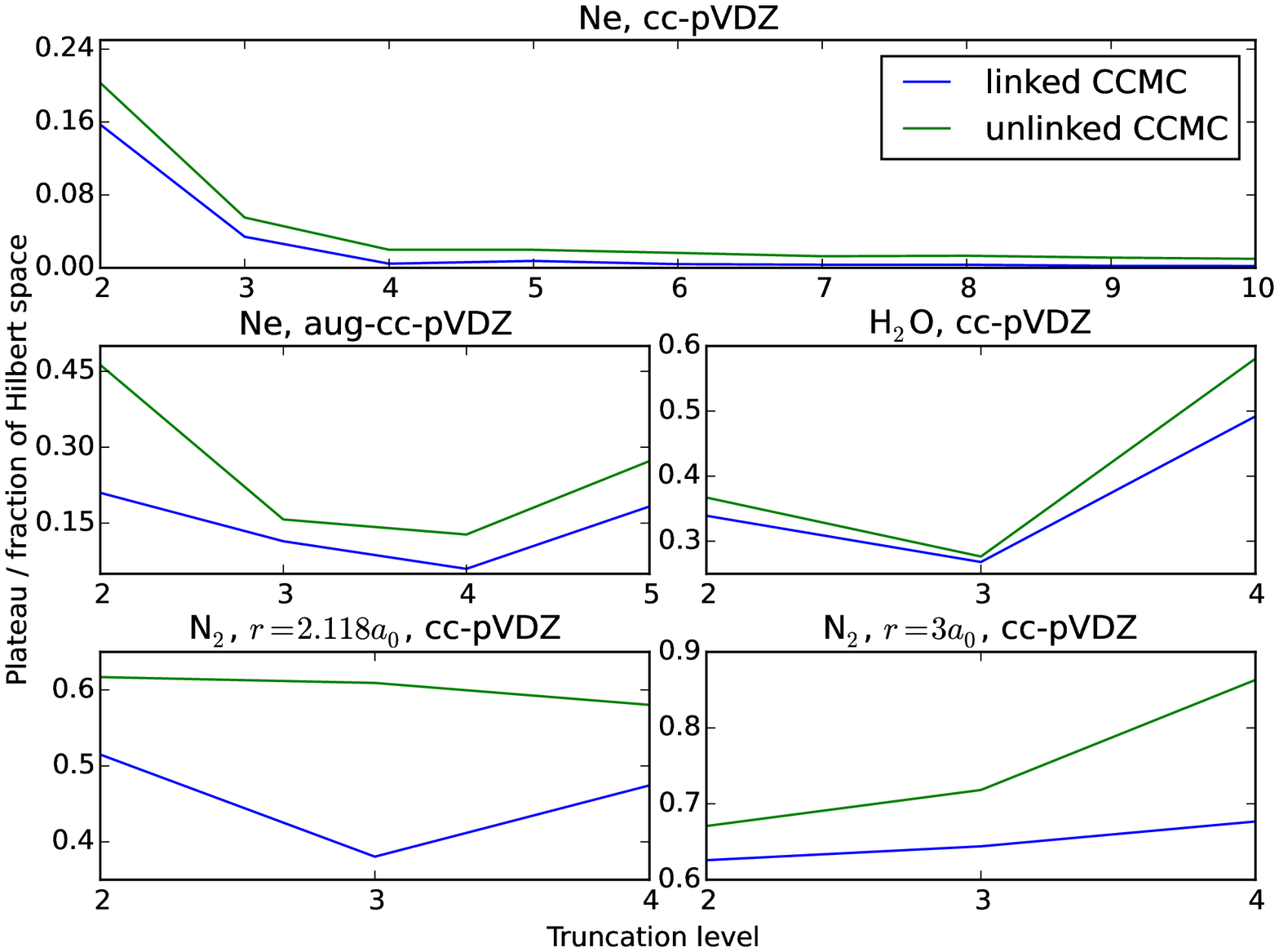}
    \caption{Plateau heights as a fraction of the size of the Hilbert space for the neon atom and nitrogen and water (geometry from Ref.~\onlinecite{Olsen96}) molecules using linked and unlinked CCMC.  In all cases, a lower plateau is obtained with the linked equations.\label{fig:plateaus} }
\end{figure}

We have studied some small systems to see the effect of this algorithm.
The coupled cluster energies are recovered with, as shown in \cref{fig:plateaus}, lower plateau heights than when using the original formulation of the algorithm.
The effect is most pronounced at high truncation levels, which is to be expected as the termination of the expansion of $\bar{H}$ at fourth order will restrict the available clusters more for higher truncations.
This means that the smaller number of connections in the effective Hamiltonian are better sampled, increasing the rate of annihilation.

Despite \cref{eq:CCMC} and \cref{eq:lCCMC} solving equivalent equations for the amplitudes $t_\mathbf{i}$, the proportion of excitors on the reference is much larger when using the linked equations.  The instantaneous populations fluctuate, so only the expectation values $\braket{ N_\mathbf{i} }$ correspond to the correct amplitudes.  When considering the total population, however, $N_\mathrm{tot} = \sum_\mathbf{i} | N_\mathbf{i} | $ is a non-linear function of $N_\mathbf{i}$, so $\braket{ N_\mathrm{tot} } \geq \sum_\mathbf{i} | \braket{ N_\mathbf{i} } |$.  Excitors on which the sign of the population is not constant raise the value of $\braket{ N_\mathrm{tot} }$.  As the linked equations reduce the number of low amplitude excitors that are instantaneously occupied at a given iteration, they give a lower $\braket{ N_\mathrm{tot} }$ for a fixed population of excitors on the reference.  We are investigating the impact of this but believe that, due to the presence of the population on the reference in the cluster generation probabilities, this results in more stable calculations.

In conclusion, we have demonstrated that this linked approach makes stochastic coupled cluster calculations more feasible, especially at the higher truncation levels, which are not typically used but are necessary for an accurate treatment of many chemically interesting problems.
We have also presented a modification to the death step for both linked and unlinked formalisms to remove a bias in projected energy estimators while the population is still growing, and we believe that this will be useful in other cases where there is non-linearity in the projector.

By placing CCMC in the same theoretical framework as conventional coupled cluster theory, namely using the similarity transformation of the Hamiltonian, we hope to be able to translate further technical and theoretical developments to our stochastic implementation.
We finish by emphasising two important benefits of a stochastic approach to coupled cluster: the substantial reduction in memory demands, by allowing sparsity in the amplitudes to emerge naturally, and access to an efficient coupled cluster algorithm that can handle arbitrary truncation levels.

All CCMC calculations were performed using a development version of HANDE\cite{Spencer14,Spencer15}, with one- and two-body molecular integrals obtained from Hartree-Fock calculations performed in Q-Chem.\cite{QCHEM}  All data was analysed using pyblock\cite{pyblock}, and plots produced using matplotlib\cite{matplotlib}.  Raw and analysed data and analysis scripts are available at Ref.~\onlinecite{linked_CC_data}.

\section*{Acknowledgements}
R.S.T.F.~is grateful to CHESS for a studentship and A.J.W.T.~to the Royal Society for a University Research Fellowship.
J.S.S.~acknowledges the research environment provided by the Thomas Young Centre under Grant No.~TYC-101. \\

\bibliography{refs}

\end{document}